%% file: main.tex
\documentclass[manuscript,screen]{acmart}

\usepackage{graphicx}
\usepackage{caption, subcaption}
\usepackage{amsmath}
\usepackage{enumitem}
\usepackage{balance}

\usepackage{xcolor}

\AtBeginDocument{%
  \providecommand\BibTeX{{%
    \normalfont B\kern-0.5em{\scshape i\kern-0.25em b}\kern-0.8em\TeX}}}

\copyrightyear{2022} 
\acmYear{2022} 
\setcopyright{rightsretained}
\acmConference[CHIWORK]{2022 Symposium on Human-Computer Interaction for Work}{June 8--9, 2022}{Durham, NH, USA}
\acmBooktitle{2022 Symposium on Human-Computer Interaction for Work (CHIWORK), June 8--9, 2022, Durham, NH, USA}
\acmPrice{15.00}
\acmDOI{10.1145/3533406.3533415}
\acmISBN{978-1-4503-9655-4/22/06}

\begin{document}

\title{The Future of Hybrid Meetings}


\author{Marios Constantinides}
\affiliation{%
  \institution{Nokia Bell Labs}
  \city{Cambridge}
  \country{United Kingdom}
}
\email{marios.constantinides@nokia-bell-labs.com}

\author{Daniele Quercia}
\affiliation{%
  \institution{Nokia Bell Labs}
  \city{Cambridge}
  \country{United Kingdom}
}
\email{daniele.quercia@nokia-bell-labs.com}

\renewcommand{\shortauthors}{Constantinides and Quercia}

\begin{abstract}
Meetings are typically considered to be the fuel of an organization's productivity---a place where employees discuss ideas and make collective decisions. However, it is no secret that meetings are also often perceived as wasteful vacuums, depleting employee morale and productivity, likely due to the fact that current technologies fall short in fully supporting physical or virtual meeting experience. In this position paper, we discuss the three key elements that make a meeting successful (i.e., execution, psychological safety, and physical comfort), and present new tools for hybrid meetings that incorporate those elements. As past research has focused on supporting meeting execution (the first element), we set the roadmap for future research on the two other elements: on psychological safety by articulating how new technologies could make meeting useful for all participants, ensure all participants give and receive appropriate levels of attention, and enable all participants to feel and make others feel comfortable; and on physical comfort by dwelling on how new technologies could make the meeting experience comfortable by integrating all human senses. We also discuss the potential danger of these technologies inadvertently becoming surveillance tools.
\end{abstract}




\maketitle

\input{sections/1.Introduction}
\input{sections/2_Proposal}

\input{sections/3_Tools}
\input{sections/4_Future}

\begin{acks}
We thank those who actively supported this research at Nokia Bell Labs; in particular, Sagar Joglekar, Bon Adriel Aseniero, and Jun-Ho Choi for their active role in the development of the meeting companion app MeetCues; and Mark Clougherty, Sean Kennedy, Michael Eggleston, and Marcus Weldon for their guidance during the development. We also thank Nigel Oseland for sharing his research at the intersection of environmental psychology and workplace strategy.
\end{acks}

\balance

\bibliographystyle{ACM-Reference-Format}
\bibliography{main}


\end{document}

%% file: sections/1.Introduction.tex
\section{Introduction}
\label{sec:introduction}
While meeting tools, to a great extent, have increasingly simplified and augmented the ways meetings are conducted, they still fall short in supporting the nuanced experience of offline meetings. To see why, consider, a virtual all-hands meeting, where one would expect a one-to-many form of communication. One could imagine that not everyone would feel comfortable sharing their video streams in such a setting and, as such, the meeting host might miss the opportunity to ``read the room'' (i.e., interpret non-verbal cues that are generally expressed in offline settings). The interplay between offline and online worlds has slowly but surely produced a new breed of meetings: \emph{the hybrid meeting}. This type of meeting typically involves a mixture of in-person and remote attendees: remote attendees join the meeting via a virtual meeting platform (e.g., Zoom), and in-person attendees sit together in the typical meeting room. Meeting experience is therefore shaped not only by the participants' interactions but also by their diverse physical environments. In this position paper, we: 
\begin{itemize}
\item Discuss the three key elements that make a meeting successful (Section~\ref{sec:proposal}). 

\item Present recent research that incorporates those three elements (Section~\ref{sec:tools}).

\item Set the roadmap for future research, focusing on the two elements out of the three that are often neglected in the literature---psychological safety and physical comfort (Section~\ref{sec:future}).

\item Discuss the potential danger of future meeting technologies inadvertently becoming surveillance tools (Section~\ref{sec:surveillance}).

\end{itemize}

%% file: sections/2_Proposal.tex
\section{What makes meetings successful}
\label{sec:proposal}

\begin{figure*}
    \centering
     \includegraphics[width=0.98\textwidth]{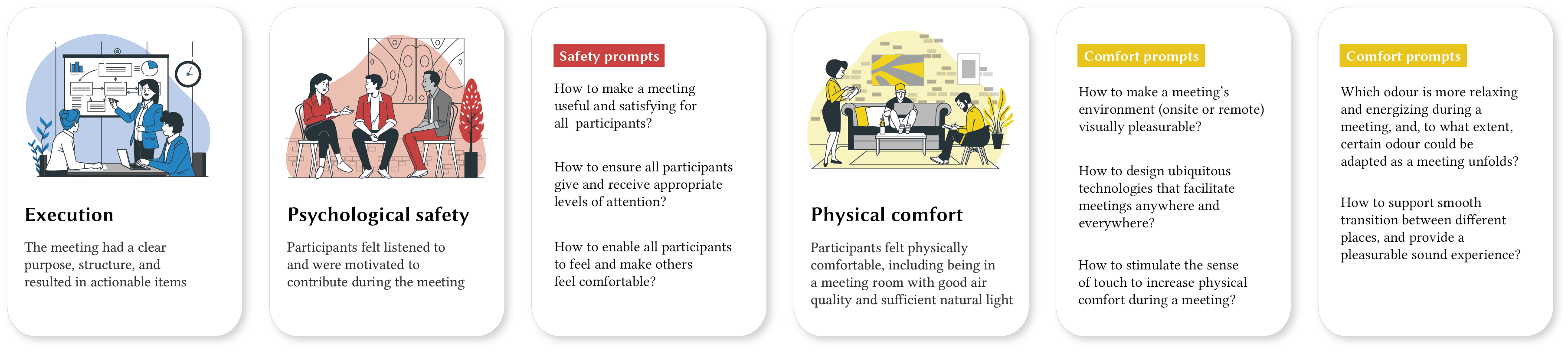}
    \caption{Three key elements that make a meeting successful along with key research questions for future research: (a) \emph{Execution}: whether the meeting had a clear purpose, structure, and resulted in actionable items; (b) \emph{Psychological safety}: whether participants felt listened to and were motivated to contribute during the meeting; and (c) \emph{Physical comfort}: whether participants felt physically comfortable by, say, being in an environment with good air quality and sufficient natural light. Most innovations in meeting technologies have focused on execution, and future work should focus on the two other aspects -  psychological safety and physical comfort. A sample of key research questions are reported in this infographic.}
    \label{fig:gaps}
\end{figure*}

A meeting's perceived experience has been evaluated through standardized questionnaires, or ad-hoc post meeting surveys. For example, the Event Performance Indices (EPI)~\cite{meetingmetrics} is a six-item questionnaire that focuses on a meeting's overall performance results, and, in turn, measures meeting effectiveness. The questionnaire prompts participants' satisfaction levels including, for example, whether the meeting was worth the time investment, and whether participants were personally motivated. In the fields of Management and Organizational Science, meeting productivity has been directly linked with a meeting's agenda, structure, and purpose~\cite{lent2015_meetings, hbr_cohen_start, hbr_schwarz_start}.
Inclusiveness~\cite{hbr_better_meetings, hbr_quality_experience}, dominance~\cite{romano2001meeting}, peripheral activities~\cite{niemantsverdriet2017recurring}, and psychological safety~\cite{hbr_tense, hbr_psychological_safety} are aspects that have also been found to influence meeting experience and productivity. The physical comfort of the workplace environment is an additional factor that makes up the list~\cite{alavi2017comfort}. For example, a recent survey found that light and outdoor views were among the most popular perks employees craved for~\cite{hbr_office_light}, whereas stuffy and stale air offices tended to reduce  productivity~\cite{allen2016associations, hbr_office_air, hbr_air_pollution}. 

More recently, to capture the factors that generally make a meeting successful, Constantinides et al.~\cite{comfeel} designed a 28-question survey and administered it to 363 individuals whose answers were then statistically analyzed. The survey covered an array of themes, previously identified in the Organization and Management Science literature, including a meeting's psychological experience (e.g., contribution, balance, turn-taking, attention), its structure, and its content (e.g., agenda, physical environment, use of technology). The survey responses were then statistically analyzed using a Principal Component Analysis, with 11 items of the 28-question survey explaining 62\% of the total variance. These 11 items loaded on three factors - the three factors that make a meeting successful:

\begin{description}
    \item \textbf{Execution}. Whether participants ultimately ``got things done''---more specifically, whether the meeting had a clear purpose, structure, and resulted in actionable items.
    \item \textbf{Psychological Safety}. Whether participants felt listened to and were motivated to contribute during the meeting.
    \item \textbf{Physical Comfort}. Whether participants felt physically comfortable by, for example, be in a meeting room with good air quality and sufficient natural light.
\end{description}

%% file: sections/3_Tools.tex
\section{Current Support For The Three Factors Determining Success}
\label{sec:tools}
Next, we discuss recent research aiming at supporting these three key elements in either physical or virtual meetings.

\subsection{Execution} Execution refers to whether the meeting had a clear purpose, structure, and resulted in actionable items. Besides the mainstream communication tools of the likes of MS Teams, WebEx, Skype, Zoom, just to name a few, a large body of scientific work has been focused on supporting meeting execution through contextual information, text, audio, and video support. For example, one work focused on detecting key decisions in dialogues~\cite{kim2014learning}, while another on generating an ``action items'' list from these dialogues~\cite{mcgregor2017more}. Using an agenda planning technique, Garcia et al.~\cite{garcia2004cutting} developed a tool that allows meeting participants to vote for agendas items. As the balance of conversational turn-taking is important for group performance~\cite{woolley2010evidence}, technologies were developed to create awareness. This was done by highlighting salient moments and visualizing participants' contributions~\cite{kim2008meeting}. Content recording was also a focus for technologies supporting execution.  NoteLook~\cite{chiu1999notelook} exploited video streams to support note taking. Catchup~\cite{tucker2010catchup} automatically identified the gist of what was missed in a meeting, allowing people to join the meeting late and still participate effectively. Video Threads~\cite{barksdale2012video} supported asynchronous video sharing for geographically distributed teams. Finally, Banerjee et al.'s  playback system allowed for revisiting a recorded meeting~\cite{banerjee2005necessity}.

\subsection{Psychological Safety}
Psychological safety refers to whether participants felt listened to and were motivated to contribute during the meeting.  As Edmondson described it, psychological safety refers to ``the absence of interpersonal fear that allows people to speak up with work-relevant content''~\cite{edmondson1999psychological}. In face-to-face interactions (e.g., during a social encounter or a meeting), we primarily read faces through visual cues and read bodies through a multi-sensory integration. All these cues, to a great extent, allow us to understand the dynamics in a conversation (e.g., whether one feels listened to and receives the appropriate attention from his/her peers). In virtual interactions, however, these natural cues are often  lost, flattening the psychological and social experience. Furthermore, current meeting technologies need to deliver this experience in an attention-enhancing way as opposed to the current attention-depleting way: currently, in a virtual meeting, attendees need to pay attention not only to what is being said, but also to peripheral modes of communication (e.g., text messages, raising of virtual hands). 

To bring the offline meeting experience into the virtual world, recent research  developed an app called MeetCues. This captures three types of cues, aiming at augmenting psychological safety in hybrid meetings. First, the MeetCues app allows meeting attendees to provide real-time feedback~\cite{meetcues}. The app enabled attendees to react on what was being discussed by tagging key points and action items. These virtual crowdsourced tags were translated into an emoji cloud visualization, which allowed attendees to infer the overall atmosphere. In this way, these crowdsourced cues served as indicators of moments during which, for example, participants agreed with what was being said or whether further clarification was needed, cultivating a safe environment for contributions. 

However, these crowdsourced feedbacks did not fully capture the multi-sensory integration that we are attuned to in physical meetings. A case in point is the almost universally acceptable non-verbal cue of nodding~\cite{peleckis2015nonverbal}. In face-to-face interactions, it is natural to observe bodily expressions to understand, for example, whether one agrees on what is being said (nodding), or a clarification is needed. However, the picture is different in virtual meetings. Participants of virtual meetings who often turn off their cameras, leave the speaker staring at a sea of black squares, feeling psychologically disoriented. 

That is why the second type of cues MeetCues captured was body movements. It did so by integrating wearable devices with the MeetCues app, which captured participants' heart rates, head and hand movements, and changes in postures~\cite{kairos, park2020wellbeat}. These body cues were predictive of the meeting's vibrancy and multi-tasking activities, and ultimately of the meeting's success; these cues were even more predictive than the meeting's emotional content derived from the meeting's transcript. For example, head movements such as nodding served as a proxy for (dis)agreement, while changes in postures served as a proxy for (dis)comfort. In particular, these two body cues metrics proved useful in helping  attendees infer the levels of psychological safety ``in the room''.

The third type of cues MeetCues captured was types of conversations~\cite{choi2020ten} (e.g., a heated discussion resulting in a conflict eventually being resolved, a supportive conversation, an exchange of knowledge). By analyzing more than four thousand minutes of conversations from eighty-five real-world meetings, Zhou et al.~\cite{zhou2021role} found that conversation types were more predictive of meeting success than traditional voice and text analytics. By monitoring these conversations during a meeting (or after it), one could potentially measure specific aspects of organizational productivity, and proactively take actions for improvement. To see how, consider conflict resolution. Having new real-time ways of marking indicators of, for example, conflicts, not only would increase attendees’ awareness but would also help cultivate an approach of conflict resolution.

\subsection{Physical Comfort}
Physical comfort refers to whether participants felt physically comfortable by, for example, be in a meeting room with good air quality and sufficient natural light. To see what we mean by the word `comfort', let us draw a parallel with architecture. In that area, comfort mainly describes four main qualities that a good building must possess: visually pleasurable, thermally livable, acoustically tolerant, and respiratory humane (in terms of air quality)~\cite{alavi2017comfort}.

To paraphrase comfort in the meeting context, recent research has initially focused on the specific aspects of good air quality and natural light. A team of researchers, in particular,  developed miniaturized devices, called Geckos, which are fitted with cheap-to-produce sensors capturing light, temperature, and the presence of a broad range of gases such as volatile organic compounds (VOCs). They integrated Geckos with the MeetCues app~\cite{meetcues}, and created a new indoor environmental sensing infrastructure ComFeel~\cite{comfeel}. At the end of each meeting, the MeetCues app asks each participant: \emph{(a)} whether the meeting had a clear purpose and structure, and resulted in a list of actionable points, and \emph{(b)} whether each participant felt listened to and was motivated to contribute. 

To explore the extent to which air quality alone determined whether a meeting was perceived to be productive or not, they deployed ComFeel in a corporate office and gathered data from 29 meetings in different rooms. As one expects, productive meetings were those in which participants felt safe to contribute: the probability of a meeting being productive increased by 35\% for each standard deviation increase in psychological safety. Surprisingly, the results for air quality were dramatic too. The productivity probability increased by as much as 25\% for each standard deviation increase in room pleasantness. Furthermore, among all the sensors, the air quality one was the most important: indeed, room pleasantness was achieved through improved temperature (with a relative contribution of 25\%), lighting (30\%), and air quality (45\%). These results suggest that, if a meeting takes place in a stuffy conference room, even if it is run well, people will still struggle to pay attention. To fix that, one just needs to do a handful of things, from manually or automatically adjusting ventilation, lighting, temperature, which could increase a meeting's productivity by a considerable extent.

%% file: sections/4_Future.tex
\section{The three factors in Future Hybrid Meetings}
\label{sec:future}

Most innovations in meeting technologies have focused on supporting physical or virtual meetings independently, and they did so by extensively focusing on how to support meeting execution. That is why future work should focus on the two other overlooked aspects (Figure~\ref{fig:gaps}): on psychological safety (Section~\S\ref{sec:futureps}) by making meetings useful for all participants, ensuring all participants give and receive appropriate levels of attention, and enabling all participants to feel and make others feel comfortable; and on physical comfort (Section~\S\ref{sec:futurepc}) by integrating all human senses, and, as such, making the meeting sensory experience pleasurable. At the same time though, future work should also consider how these new technologies could inadvertently become surveillance tools (Section~\S\ref{sec:surveillance}). 

\subsection{Rethinking Psychological Safety}
\label{sec:futureps}
The design space of meeting technologies could be well enriched with features that fully support all three facets of psychological safety~\cite{comfeel}: \emph{(a)} making the meeting useful and satisfying for all participants, \emph{(b)} ensuring all participants give and receive appropriate levels of attention, and \emph{(c)} enabling all participants to feel and make others feel comfortable. 

First, to make a meeting useful and satisfying, existing tools already provide analytics in the form of summaries or real-time feedback~\cite{meetcues}. However, these analytics could be enriched with aspects that are hard to quantify such as the types of conversations happening in a meeting~\cite{zhou2021role}, nuanced emotional states~\cite{zhou2022predicting}, or other language markers that are hidden in conversations (e.g., presence of stress, engagement in emphatic conversations)~\cite{scepanovic2020extracting, robertson2019language, zhou2021language}. 

Second, as turn-taking and balance in conversation helps to cultivate a safe environment for contribution~\cite{woolley2010evidence}, future meeting tools need to ensure that all attendees give and receive the appropriate levels of attention. For example, some meeting technologies already do provide indicators about ``speaking up'' time, and often do so through attention-depleting notifications. However, future technologies must ensure attention-enhancing ways of notifying users. As Weiser and Brown argued in their seminar work \emph{The Coming Age of Calm Technology}, ``the information display should move easily from the periphery of users’ attention to the centre, and back''~\cite{weiser1997coming}. Future design elements need to smoothly capture the user's attention only when necessary, while calmly remain in the user's periphery most of the time. For example, while current meeting tools allow various modes of communication (e.g., voice, text) and notifications (e.g., emojis) in a virtual meeting, personalized or timely delivery of those notifications could remove unnecessary information overload. Imagine a scenario during which five attendees need to focus on the speaker, yet one of them types a question and another raises his/her virtual hand. If these notifications were to be delivered straight away, they may well overload the speaker and distract the audience. More work needs to go into  identifying the right moment for delivering a notification, or that for delivering a notification to the best person in a group, allowing for personalized delivery.

Third, it is also important for meeting technologies to facilitate an inclusive environment where all attendees feel comfortable, and can make others feel comfortable. One way of achieving it is to create social interactions that scaffold learning, and then use computational methods to weave these interactions into the fabric of meeting tools. For example, as demonstrated in massive online classes (MOOCs)~\cite{quintana2018mentor}, future meeting technologies could coach people skills that foster inclusion and diversity, and collaboratively help them reflect on their behavior. Another way of achieving inclusivity is through broadening the design space. Design elements such as emojis allow, to a great extent, attendees to non-verbally communicate and express their emotional state; it can be seen as a way of developing trust or empathy for others. Future meeting technologies could also borrow concepts from biophilic design~\cite{wilson1984biophilia} and embrace new types of visual cues~\cite{qin2020heartbees} (e.g., the use of different symbols, imagery, and artificial artifacts), thus making additional layers of non-verbal communication available to users.

\subsection{Rethinking Physical Comfort}
\label{sec:futurepc}
Aligned with the Human-Building Interaction vision~\cite{alavi2019introduction} (an emerging area aiming at unifying HCI research in the built environment), physical comfort goes well beyond air quality, integrating all the human senses. To begin with, let's take sight. Could we make the physical environment visually pleasurable? Recent studies showed that adjusting light conditions in a room could serve as a stress reducing intervention mechanism, or even as a biofeedback relaxation training~\cite{ren2019lightsit, yu2018delight}. Moving to hearing. Could we design ubiquitous technologies that facilitate meetings anywhere and everywhere? Consider a knowledge worker who attends meetings from a variety of places: from the office to home to a cafeteria to even an autonomous car~\cite{kun2020future}. How could future technologies support smooth transition between different places, and provide a pleasurable sound experience? Smart earable devices can be one of the answers wherein an array of sensors could be embedded in them to control not only what sound the end user is hearing, but also how that sound is delivered in a specific physical setting~\cite{kawsar2018earables}. While addressing sight and hearing is to some extent easier, smell and touch require future research efforts. Which odour is more relaxing and energizing during a meeting, and, to what extent, certain odour could be adapted as a meeting unfolds to facilitate better execution? How could we stimulate the sense of touch to increase physical comfort during a meeting? For example, as most of white collar jobs require employees to spend long hours sitting, previous studies explored the design of ergonomic chairs that would increase physical comfort levels. While these studies typically employ various objective methods to capture physical comfort such as pressure sensors measurement~\cite{zemp2015pressure} or an array of sensors including, for example, temperature,
gas, illuminance, VOC, CO\textsubscript{2}~\cite{zhong2020hilo}, new sensing modalities such as Electromyography (EMG) could prove useful to fully incorporate the sense of touch in the design space of these chairs. 

\section{The dangers of meeting technologies}
\label{sec:surveillance}
While these technologies hold the promise of enabling employees to be productive, report after report has highlighted the outcries of workplace AI-based solutions being biased and unfair, and lacking transparency and accountability. During the COVID-19 pandemic, systems were being used to analyze footage from security cameras in workplace to detect when employees are not complying with social distancing rules\footnote{\url{https://www.ft.com/content/58bdc9cd-18cc-44f6-bc9b-8ca4ac598fc8}}; while there is a handful of good intentions behind such a technology, the very same technology could be used for tracking employees' movements, or time away from desk. As we move towards a future likely ruled by big data and powerful AI algorithms, important questions arise relating to the psychological impacts of surveillance, data governance, leadership and organizational culture, and compliance with ethical and moral concerns. The historian and philosopher Yuval Noah Harari argued that digital platforms such as those that allow us to work from remote need to follow three basic rules\footnote{\url{https://www.ft.com/content/f1b30f2c-84aa-4595-84f2-7816796d6841}} to protect us from ``digital dictatorships''. First, any data collection on people should be used to help people rather than to manipulate, control, or harm them. In the meetings context, this translates into providing analytics that help employees reflect on their experience rather than causing them to receive low performance review in the event of a meeting not being executed well. Second, surveillance must always go both ways. That is, whenever an organization increases surveillance of individuals, at the same time, organizational accountability needs to increase. If organizations could establish processes to monitor their workforce, they may well establish processes to audit their own actions as well. Third, data should not be concentrated in a single entity, not least because data monopolies are the recipe for dictatorship. In the workplace context, this translates into newly created divisions in a company that oversee data collection and, ideally, ensure that data about their workers is in the hands of the workers themselves.

In a global modernized workplace, new meeting tools are likely to be developed, facilitating and augmenting the experience of hybrid meetings. As a result of the COVID-19 pandemic, fully remote or hybrid meetings removed any physical barriers and often contributed to improved work-life balance~\cite{rudnicka2020eworklife}. On the downside, tools for supporting hybrid meetings may inadvertently becoming surveillance tools, compromising employees' privacy~\cite{surveillance}. To unpack the AI ethics of workplace technologies, Constantinides and Quercia~\cite{surveillance} conducted a crowdsourcing study to understand how employees judge such technologies and determine which ones are desirable, and why. They considered 16 workplace technologies that track productivity based on diverse inputs (e.g., tracking audio conversation during virtual meetings, tracking text messages in collaboration tools), and asked crowd-workers to judge these scenarios along five moral dimensions. They found that workplace technologies were judged harshly depending on three aspects (heuristics) participants used to assess the scenarios. In increasing importance, these aspects reflected whether a scenario: 1) was not currently supported by existing technologies (\emph{hard to adopt}); 2) interfered with current ways of working (\emph{intrusive}); and, more importantly, 3) was not fit for tracking productivity or infringed on individual rights (\emph{harmful}). Tracking eye movements in virtual meetings and the visited websites in remote work, despite being possible, were considered to be ``on the way'' of getting the job done (they were easy to adopt but intrusive). By contrast, tracking text messages in collaboration tools such as Slack was considered to not interfere with work (unobtrusive). Finally, tracking audio conversations in virtual meetings was considered to be possible (easy to adopt), and not interfere with work (unobtrusive), yet it was considered to be harmful, as it entailed tracking not only whether a meeting took place but also its content, causing a loss of control. 
The above heuristics offer a guide on how workplace technologies are likely to be morally judged. Having a technology that is easy to implement and does not interfere with work is not necessarily a technology that should be deployed. \emph{Tracking facial expressions} (even beyond the nefarious uses - of dubious effectiveness - of inferring political orientation or sexual preferences~\cite{wang2018deep}) is possible and could be done in seamless ways (e.g., with existing off-the-shelf cameras), yet it would be still considered harmful and unethical. \emph{Tracking eye movements}, \emph{task completion}, or \emph{typing behavior} was considered a proxy for focus (harmless) yet intrusive as it would ``get in the way''. \emph{Tracking social media use in remote work} was considered not only intrusive but also harmful, as it infringes on privacy rights.

On a very pragmatic level, there is a handful of reasons as to why organizations opt for employee surveillance (e.g., maintaining productivity, monitoring resources used, protecting the organization from legal liabilities). Critics, however, rightly argue that there is a fine line between what organizations could be monitoring and what they should be monitoring. If this line is crossed, it will have consequences on employees, affecting their well-being, work culture, and productivity~\cite{ball2010workplace}. If future meeting tools incorporate any kind of employee monitoring, they need to preserve individual rights, including that of privacy. New meetings tools need to also ensure non-discrimination, while promoting inclusivity, and, at the same time, provide explainable and understandable outputs (e.g., the decision upon which the system decided that a meeting was successful or not).

Hybrid meetings also create asymmetries of interactions stemming from the social and cultural contexts. As Saatci et al.\cite{saatcci2019hybrid} found, the most challenging asymmetry is the diverse experience between co-located and remote meeting participants. Remote participants often feel isolated, while co-located participants dominate the interaction. Differences in language and accent, cultural behaviors, digital literacy, physical location, and the boundaries between work and personal life~\cite{rudnicka2020eworklife} also contribute to interaction asymmetries. To overcome such challenges, new meetings tools should focus on making meetings more inclusive for everyone by maximizing psychological safety and optimizing physical comfort.